\documentclass[floats,twocolumn,aps,prb,longbibliography,superscriptaddress]{revtex4-2}

\usepackage{graphicx}
\usepackage{amsfonts}
\usepackage{amssymb}
\usepackage{amsbsy}
\usepackage{amsmath}\usepackage[colorlinks = true,
            linkcolor = magenta,
            urlcolor  = blue,
            citecolor = blue,
            anchorcolor = magenta]{hyperref}
\usepackage{hyperref}
\usepackage{xcolor}
\usepackage{subeqnarray}
\usepackage{soul}
\usepackage[version=4]{mhchem}

\allowdisplaybreaks[1]

\begin{document}




\title{Dynamics of K$_2$Ni$_2$(SO$_4$)$_3$ governed by proximity to a 3D spin liquid model}

\author{M. G. Gonzalez}
\affiliation{Helmholtz-Zentrum Berlin f\"ur Materialien und Energie, Hahn-Meitner-Platz 1, 14109 Berlin, Germany}
\affiliation{Dahlem Center for Complex Quantum Systems and Fachbereich Physik, Freie Universität Berlin, 14195 Berlin, Germany\looseness=-1}

\author{V. Noculak}
\affiliation{Helmholtz-Zentrum Berlin f\"ur Materialien und Energie, Hahn-Meitner-Platz 1, 14109 Berlin, Germany}
\affiliation{Dahlem Center for Complex Quantum Systems and Fachbereich Physik, Freie Universität Berlin, 14195 Berlin, Germany\looseness=-1}

\author{A. Sharma}
\affiliation{Laboratory for Quantum Magnetism, Institute of Physics, \'Ecole Polytechnique F\'ed\'erale de Lausanne, CH-1015 Lausanne, Switzerland\looseness=-1}

\author{V. Favre}
\affiliation{Laboratory for Quantum Magnetism, Institute of Physics, \'Ecole Polytechnique F\'ed\'erale de Lausanne, CH-1015 Lausanne, Switzerland\looseness=-1}

\author{J-R. Soh}
\affiliation{Laboratory for Quantum Magnetism, Institute of Physics, \'Ecole Polytechnique F\'ed\'erale de Lausanne, CH-1015 Lausanne, Switzerland\looseness=-1}

\author{A. Magrez}
\affiliation{Crystal Growth Facility, \'Ecole Polytechnique Fédérale de Lausanne, Lausanne, Switzerland.\looseness=-1}

\author{R. Bewley}
\affiliation{ISIS Pulsed Neutron and Muon Source, STFC Rutherford Appleton Laboratory, Harwell Science and Innovation Campus, Didcot, Oxfordshire, OX11 0QX UK\looseness=-1}

\author{Harald O. Jeschke}
\affiliation{Research Institute for Interdisciplinary Science, Okayama University, Okayama 700-8530, Japan}
\affiliation{Department of Physics and Quantum Centre of Excellence for Diamond and Emergent Materials (QuCenDiEM), Indian Institute of Technology Madras, Chennai 600036, India}

\author{J. Reuther}
\affiliation{Helmholtz-Zentrum Berlin f\"ur Materialien und Energie, Hahn-Meitner-Platz 1, 14109 Berlin, Germany}
\affiliation{Dahlem Center for Complex Quantum Systems and Fachbereich Physik, Freie Universität Berlin, 14195 Berlin, Germany\looseness=-1}
\affiliation{Department of Physics and Quantum Centre of Excellence for Diamond and Emergent Materials (QuCenDiEM), Indian Institute of Technology Madras, Chennai 600036, India}

\author{H. M. R{\o}nnow}
\affiliation{Laboratory for Quantum Magnetism, Institute of Physics, \'Ecole Polytechnique F\'ed\'erale de Lausanne, CH-1015 Lausanne, Switzerland\looseness=-1}

\author{Y. Iqbal}
\affiliation{Department of Physics and Quantum Centre of Excellence for Diamond and Emergent Materials (QuCenDiEM), Indian Institute of Technology Madras, Chennai 600036, India}

\author{I. \v{Z}ivkovi\'c}
\affiliation{Laboratory for Quantum Magnetism, Institute of Physics, \'Ecole Polytechnique F\'ed\'erale de Lausanne, CH-1015 Lausanne, Switzerland\looseness=-1}


\date{\today}

\maketitle

\textbf{Quantum spin liquids (QSLs) have become a key area of research in magnetism due to their remarkable properties, such as long-range entanglement, fractional excitations, pinch-point singularities, and topologically protected phenomena. In recent years, the search for QSLs has expanded into the three-dimensional world, where promising features have been found in materials that form pyrochlore and hyper-kagome lattices, despite the suppression of quantum fluctuations due to high dimensionality. One such material is the $S = 1$ K$_2$Ni$_2$(SO$_4$)$_3$ compound, which belongs to the langbeinite family consisting of two interconnected trillium lattices. Although magnetically ordered, K$_2$Ni$_2$(SO$_4$)$_3$ has been found to exhibit a highly dynamical and correlated state which can be driven into a pure quantum spin liquid under magnetic fields of only $B \simeq 4$~T. In this article, we combine inelastic neutron scattering measurements with pseudo-fermion functional renormalization group (PFFRG) and classical Monte Carlo (cMC) calculations to study the magnetic properties of K$_2$Ni$_2$(SO$_4$)$_3$, revealing a high level of agreement between the experiment and theory. We further reveal the origin of the dynamical state in K$_2$Ni$_2$(SO$_4$)$_3$ by studying a larger set of exchange parameters, uncovering an `island of liquidity' around a focal point given by a magnetic network composed of tetrahedra on a trillium lattice.}

QSLs are highly-entangled states of matter, in which no long-range magnetic order is observed even in the absence of thermal fluctuations at zero temperature. Ever since P. W. Anderson's proposal of a resonating valence bond phase as the ground state for the triangular lattice Heisenberg antiferromagnet~\cite{Anderson73}, QSLs have captured the attention of physicists across fields beyond quantum magnetism~\cite{Savary17}. The reason lies in the wide range of exotic properties and phenomena that these intriguing states of matter display, ranging from fractionalization of spin excitations observed as an extended continuum in the excitation spectrum, to pinch-point singularities observed in the static spin-spin correlations and associated with U(1) QSLs or fracton phases~\cite{Henley05, Fennell09, Benton16, Prem18, Niggemann23}.

The QSL behaviour is driven by strong zero-point quantum fluctuations, which are enhanced in the presence of high magnetic frustration. This is realized either by competing isotropic interactions, as in the Heisenberg model on the kagome lattice~\cite{Sachdev92, Lecheminant97, Mila98}, or by anisotropic interactions such as in the Kitaev model on the honeycomb lattice~\cite{Kitaev06, Jackeli09, Takagi19}. On the other hand, low dimensionality also amplifies quantum fluctuations, which are thus more noticeable in one- or two-dimensional systems. However, in recent years much attention has been put into three-dimensional (3D) models and compounds. Even though zero-point quantum fluctuations are greatly suppressed in 3D, highly-frustrated systems like the network of corner-sharing tetrahedra realized by the pyrochlore lattice still provide a suitable environment for the existence of QSLs, both theoretically and experimentally~\cite{Gingras14, Gao19, Chern22}. Indeed, several compounds have been synthesized which realize pyrochlore, hyperkagome, or hyper-hyperkagome lattices and display QSL behavior in 3D~\cite{Okamoto07, Chillal20}.


\begin{figure*}[!t]
\centering
\includegraphics*[width=0.8\textwidth]{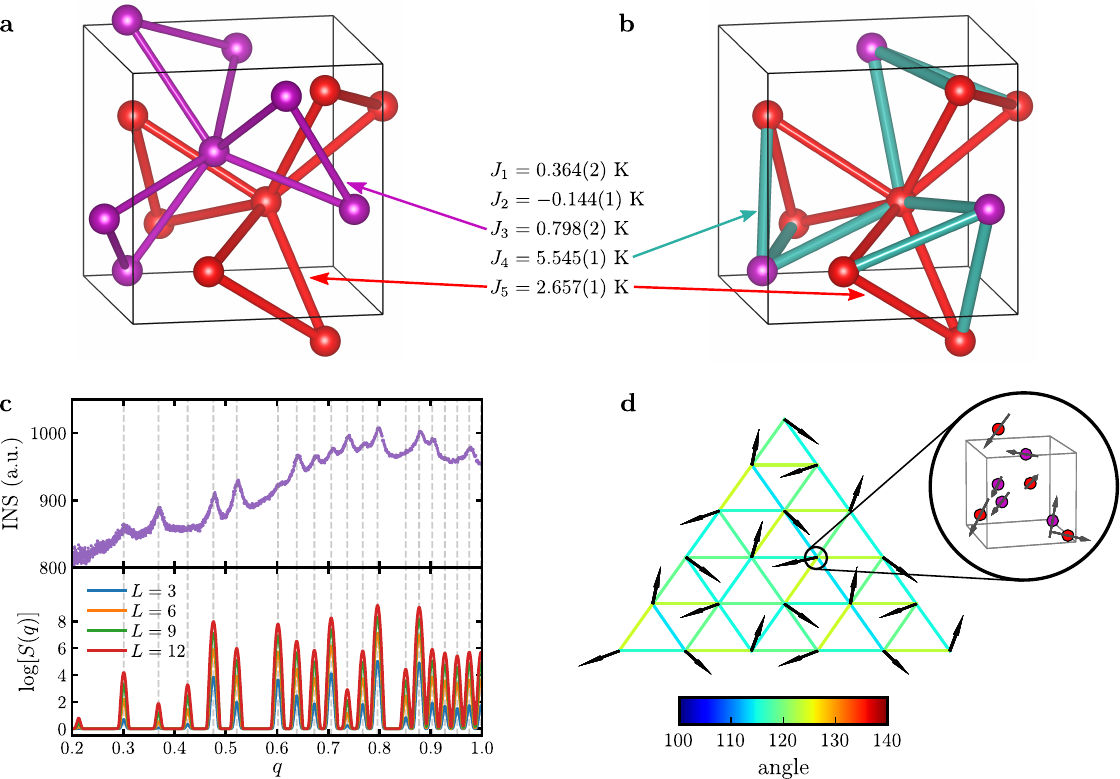}
\caption{\textbf{Crystal and magnetic structure of K$_2$Ni$_2$(SO$_4$)$_3$.} (a) Two trillium lattices of Ni$^{2+}$ ions in K$_2$Ni$_2$(SO$_4$)$_3$. (b) $J_4$ couples each ion from one trillium lattice to the nearest triangle of the second trillium lattice. For $J_4 = J_5$ magnetic ions form a network of corner-shared tetrahedra based on a trillium lattice, a tetra-trillium lattice. (c) Spin structure factor as a function of $q = |\mathbf{q}|$. Comparison between cMC calculations at $T=0$ for the DFT model (bottom) and INS measurements at 100~mK taken from Ref.~\citep{Zivkovic21} (top). A small Gaussian broadening is used for the cMC results. (d) The spin structure corresponding to the tripled magnetic unit cell determined by cMC is shown in a cut along the (111) plane for a $L=6$ system. Black arrows represent the direction of the sum of moments within a single unit cell. The bond color indicates the angle between the unit cell moments.}
\label{fig:structure}
\end{figure*}


Recently, it has been shown that a new 3D magnetic network exhibits a highly dynamical ground state. K$_2$Ni$_2$(SO$_4$)$_3$, a member of the langbeinite family, develops spin correlations below 20\,K between $S = 1$ moments, with a peculiar ordered state arising below $\sim 1.1$\,K, freezing only around 1\% of the available magnetic entropy and showing a tripling of the magnetic unit cell~\cite{Zivkovic21,Yao23}. Furthermore, the application of $B = 4$\,T magnetic field leads to a fully dynamical state down to the lowest temperatures. Another member of the langbeinite family, the KSrFe$_2$(PO$_4$)$_3$ compound with $S = 5/2$, was also claimed to exhibit spin-liquid behavior~\cite{Boya2022}.

The underlying magnetic network in these compounds comprises two interconnected trillium lattices, based on two symmetry-inequivalent magnetic sites, shown in Figure~\ref{fig:structure}a. Each trillium lattice with four sites per unit cell consists of a network of corner-sharing triangles where each site participates in three triangles. For K$_2$Ni$_2$(SO$_4$)$_3$, density functional theory (DFT) based energy mapping revealed three dominant exchange interactions, $J_3$ and $J_5$ constituting nearest-neighbour couplings within each individual trillium lattice, and $J_4$ which couples the two trillium lattices (see the full list in Figure~\ref{fig:structure}b). In the limit $J_3,J_5 \rightarrow 0$, the network transforms into a bipartite lattice, supporting a semi-classical AFM state. The limit $J_4 \rightarrow 0$ describes two independent trillium lattices, which have been theoretically investigated in detail~\cite{Hopkinson06,Isakov08} and shown to lead to a variant of the 120$^\circ$ order. Therefore, K$_2$Ni$_2$(SO$_4$)$_3$, with its magnetically highly correlated and dynamical state, represents a surprising revelation which indicates the proximity to an `island of liquidity', which has evaded the attention of the scientific community so far.


\begin{figure*}[!t]
\centering
\includegraphics*[width=0.95\textwidth]{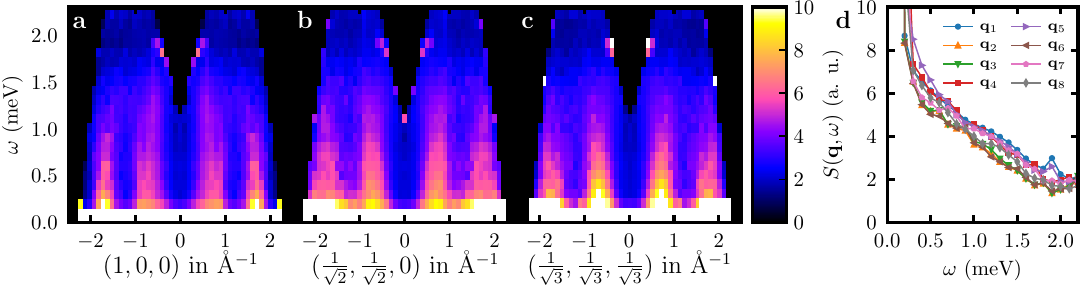}
\caption{\textbf{Time-of-flight neutron scattering on single crystals of K$_2$Ni$_2$(SO$_4$)$_3$.} Energy versus \textit{q}-vector for (a) \textbf{q} = (100), (b) \textbf{q} = (110) and (c) \textbf{q} = (111) directions. (d) Energy dependence of the dynamical structure factor obtained at several {\bf q}-points (for which streaks of intensity are observed), revealing a monotonous decrease of intensity towards the high-energy background. {\bf q}-points are: \bf{q}$_1=(0.50,0.50,0.00)$, \bf{q}$_2=(0.00,1.70,0.00)$, \bf{q}$_3=(0.65,0.00,0.00)$, \bf{q}$_4=(1.20,1.35,0.00)$, \bf{q}$_5=(0.50,0.50,0.50)$, \bf{q}$_6=(1.70,0.00,0.00)$, \bf{q}$_7=(0.00,1.65,1.65)$, \bf{q}$_8=(1.05,1.40,1.40)$.}
\label{fig:EvsQ}
\end{figure*}


Before addressing the dynamic part of the system, we focus briefly on the peculiarities of the magnetic order and how it changes depending on the different sets of exchange parameters $J_1$-$J_5$ proposed for K$_2$Ni$_2$(SO$_4$)$_3$. While previous coupling values reported corresponded to the structure at room temperature~\cite{Zivkovic21}, here we propose a new set of parameters calculated by DFT energy mapping based on a refined crystal structure at $T=100$\,K (see SI). The new values of exchange interactions (listed in Figure~\ref{fig:structure}) are moderately renormalized compared to room temperature values but show the same hierarchy of interactions $J_4 > J_5 > J_3 > J_1 > |J_2|$, where $J_2$ is the only ferromagnetic coupling. Specifically, room temperature couplings relative to the dominant $J_4$ are  $J_1=0.079J_4$, $J_2=-0.030J_4$, $J_3=0.203J_4$, $J_5=0.472J_4$, and at
$T=100$\,K they become $J_1=0.066J_4$, $J_2=-0.026J_4$, $J_3=0.144J_4$, $J_5=0.479J_4$. The most significant difference occurs for $J_3$, which changes from being $20.3\%$ of $J_4$ to $14.4\%$ of $J_4$. We performed classical Monte Carlo (cMC) calculations using the new set of exchange interactions for several system sizes $N=8L^3$, where $L^3$ is the number of unit cells and values of $L$ are taken up to 12. All calculations indicate a transition to a magnetically ordered phase. However, the lowest ground-state energy is only reached for $L = 3n$ lattices, with a critical temperature of $T_c^\mathrm{cMC}/J_4 = 0.048(2)$ evidencing the highly-frustrated nature of the Hamiltonian. Reassuringly, the obtained magnetic orders for $L = 12$, 9, 6, and 3 are all identical, indicating a tripling of the magnetic unit cell, with Bragg points in perfect agreement with experimental observations (Figure~\ref{fig:structure}c). This agreement is remarkable given that in K$_2$Ni$_2$(SO$_4$)$_3$ quantum fluctuations strongly reduce the ordered moment while our classical calculations simulate a fully static ground state. All other lattice sizes ($L\neq3n$) give higher energies at $T = 0$, implying that the magnetic order is frustrated by periodic boundary conditions. Interestingly, when the DFT Hamiltonian corresponding to the room temperature structure is used, $L = 4n$ magnetic order has the lowest energy (only 0.09\% below $L = 3n$), indicating a very sensitive landscape of complex magnetic configurations. Another recently suggested set of values with $J_3 = 0$~\cite{Yao23} leads to a quintupling of the magnetic unit cell, exhibiting the lowest ground-state energies for $L=5n$ lattices. The calculated magnetic structure for $L=3n$, shown in Figure~\ref{fig:structure}d, comprises several \textbf{q} vectors and results in a rather complicated pattern, devoid of any particular spin textures. In Figure~\ref{fig:structure}d, we show a view along the (111) plane for a $L=6$ system, where black arrows on each node represent the direction of the net magnetic moment within each unit cell. The angles between neighbouring cells are found to be close to 120$^\circ$.

To characterize the dynamical features in K$_2$Ni$_2$(SO$_4$)$_3$, we have conducted inelastic neutron scattering (INS) experiments, within the (\textit{HLL}) scattering plane. In Figure~\ref{fig:EvsQ}, we present experimental results of INS with the incident energy $E_i = 2.8$\,meV obtained at 2\,K, just above the appearance of ordering but well within the dynamical state. Energy-momentum plots along three principal directions show streaks of intensity without any apparent {\bf q}-dependence, indicating a non-dispersive type of excitations. The upper energy bound of excitations is found to be around 2\,meV, in agreement with the onset of correlations in specific heat below 20\,K~\cite{Zivkovic21}. The energy dependence of the intensity of the scattering signal for various {\bf q}-points (that present streaks) is featureless, with approximately linear decrease with energy (see panel (d) in Figure~\ref{fig:EvsQ}). Such $Q$-and $E$-dependence of the scattering intensity $S(\textbf{q},\omega)$ allows us to estimate the equal-time magnetic structure factor
\begin{equation}
    S_{\rm exp}(\textbf{q}) \sim \int_{\omega_1}^{\omega_2} S_{\rm exp}(\textbf{q},\omega) \,d\omega
    \label{eq:Sqw}
\end{equation}
which can then be directly compared with theoretical predictions based on a given spin Hamiltonian.

The magnetic structure factor $S(\textbf{Q})$ in semiclassically ordered states displays sharp and well-defined peaks, alongside magnonic excitations in the dynamic structure factor. At $E = 0$, the latter originate from the Goldstone modes that recover the broken symmetry of the magnetic phase. On the other hand, quantum spin liquids display more spread and diffuse features in the static part, where clear peaks are absent and the dynamic part is dominated by continua corresponding to fractional excitations~\cite{Paddison17, Banerjee17}. In many cases, however, features from both limits such as magnon branches and fractionalization coexist~\cite{DallaPiazza15, Ma16, Ito17, Ghioldi18}. It is worth emphasizing that fluctuations in spin liquids are far from being just incoherent and trivial disorder, and often generate specific patterns that can be associated with the characteristics of spin liquids in question, like pinch points observed in Ho$_2$Ti$_2$O$_7$, Dy$_2$Ti$_2$O$_7$, and Nd$_2$Zr$_2$O$_7$ spin ices~\cite{Fennell09, Morris09, Petit16}, or bow ties, half moons, and other theoretically predicted structures~\cite{Yan18, Mizoguchi18, Kiese23}.

\begin{figure*}[!t]
\centering
\includegraphics[width=0.95\textwidth]{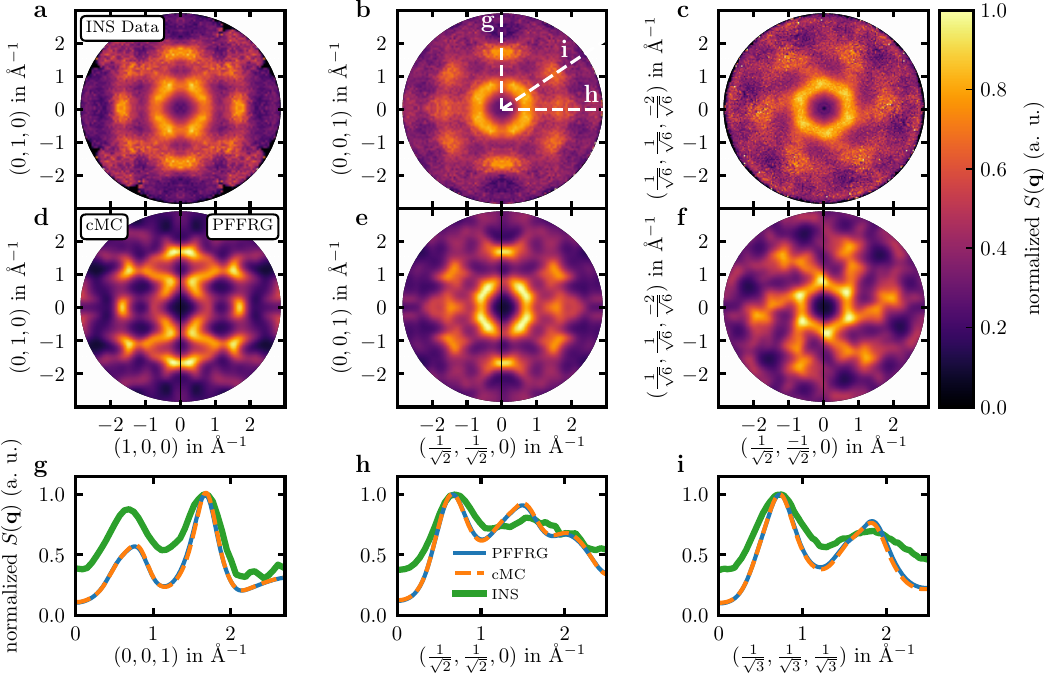}
\caption{\textbf{Comparison of experimental and theoretical spin structure factors.} Spin structure factor along different planes in reciprocal space. (a)-(c) correspond to the experimental data obtained by INS with the incident energy $E_i = 5.0$\,meV, integrated in the range 0.5\,meV and 1.0\,meV. (d)-(f) are results of cMC calculations at $T=0.35~J_4$ (left half) and PFFRG calculations for $\Lambda = 0.58~J_4$ (right half), using the form factor of Ni$^{2+}$ ions. (g)-(i) display line cuts along three principal directions indicated by white dashed lines in the panel (b).}
\label{fig:ssfplanes}
\end{figure*}

In the top row of Fig.~\ref{fig:ssfplanes} we show $S_{\rm exp}(\textbf{q})$ along three different planes in reciprocal space, obtained by integrating the intensity in the energy range $\omega_1\sim E_{\rm min} = 0.5$\,meV and $\omega_2\sim E_{\rm max} = 1$\,meV. The patterns depicted carry fingerprints from the underlying magnetic lattice composed of two interconnected trillium lattices, with a hexagonal galaxy-like motif reflecting the chirality of the crystal structure (see panel (c)), arising from the non-centrosymmetric space group.


To reproduce these experimental patterns, we performed calculations using both quantum and classical approaches. In the quantum limit, we use Pseudo-Fermion Functional Renormalization Group (PFFRG) which has been shown to produce accurate results in highly-frustrated systems~\cite{Hering22, Hagymasi22, Noculak23, Kiese23, muller2023pseudofermion}. Within PFFRG, the renormalization-group flow breaks at $\Lambda = 0.582~J_4$, indicating the presence of a magnetically ordered ground state. Static spin correlations are measured at this point and, even though calculations are carried out at $T = 0$, the finite value of the renormalization group parameter $\Lambda$ produces an effect similar to finite temperature~\cite{Iqbal-2016}. On the other hand, cMC calculations enter a classically ordered phase at low temperatures, which leads to the appearance of Bragg peaks. Thus, in an attempt to simulate the experimental data within a classical approach, we perform cMC calculations above the finite-temperature phase transition. Particularly, for $T = 0.35~J_4$, we find an excellent agreement with the PFFRG calculations (see left and right panels separated by black lines in panels (d)-(f) of Figure~\ref{fig:ssfplanes}). The striking resemblance between quantum and classical correlations at different temperatures has already been studied in Heisenberg models under the name quantum-to-classical correspondence~\cite{Tao20}. In Ref.~\cite{Tao20}, the authors focused on a mapping of exchange interactions and temperature between the quantum and classical models to obtain a quantitatively accurate correspondence. In our present case, we only allowed temperature in cMC calculations to be varied, with exchange parameters kept fixed. This quantum-to-classical correspondence indicates a mapping between the quantum $S=1$ calculations at finite renormalization group parameter $\Lambda$ and classical calculations at finite temperature $T$ and hints that quantum fluctuations can be well reproduced by thermal fluctuations in this case.


\begin{figure*}[!t]
\centering
\includegraphics[width=0.95\textwidth]{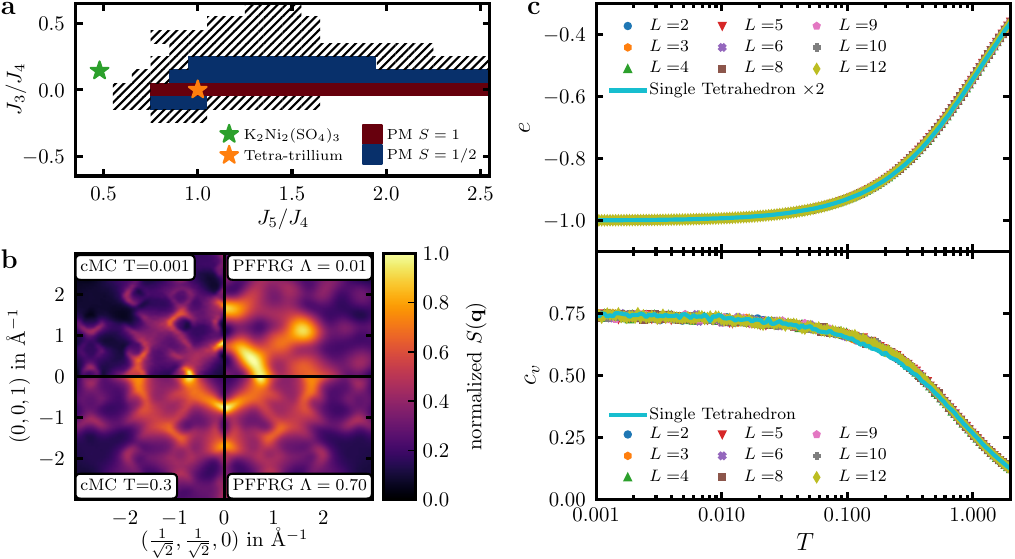}
\caption{\textbf{The 'island of liquidity' around the tetra-trillium lattice.} (a) PFFRG paramagnetic region for $S=1/2$ (red+blue) and $S=1$ (red). The dashed part indicates the region where the existence of a flow breakdown is hard to determine for the $S=1/2$ model. Green and orange stars indicate the DFT model for K$_2$Ni$_2$(SO$_4$)$_3$ and the tetra-trillium lattice limit, respectively. (b) cMC and PFFRG calculations for the tetra-trillium lattice for different values of $T$ and $\Lambda$. (c) cMC calculations for the tetra-trillium lattice limit show no finite-size effects in the energy and specific heat. Data for a single tetrahedron is also shown in light-blue lines.}
\label{fig:TT}
\end{figure*}


A more detailed comparison is obtained along the line cuts shown in panels (g-i) of Figure~\ref{fig:ssfplanes}. As is evident from all three plots, there is very little difference between cMC and PFFRG results, upholding the quantum-to-classical correspondence. The agreement with the experiments is excellent in terms of the determination of peaks in the spin structure factor, which gives rise to pattern matching in the colour plots. Deviations can be seen in terms of the predicted ratio of intensities of two principal peaks along [100] and [111], the width of peaks, as well as a qualitative difference at high \textit{q} along [110]. At least partially, these deviations could be explained by the limitations of the finite energy range integration used in Eq.~\ref{eq:Sqw}. Overall, however, it can be assumed that the theoretical description of K$_2$Ni$_2$(SO$_4$)$_3$ captures all the main aspects of its magnetism, covering both static and dynamic elements. This encourages us to investigate why its magnetic state deviates so profoundly from semi-classical AFM orders found in a bipartite lattice ($J_3, J_5 \rightarrow 0$) and in a single trillium lattice ($J_4 \rightarrow 0$).

To this end, we have studied a wider range of the $J_3$-$J_4$-$J_5$ parameter phase space and looked for signatures of static order in both PFFRG and cMC calculations. It turns out that the set of exchange parameters characterizing K$_2$Ni$_2$(SO$_4$)$_3$ lies very close to a so-far unexplored region of genuine spin-liquid behaviour, centred around a particular point defined by $J_3 = 0$, $J_4 = J_5$. In this limit, the lattice is composed of corner-sharing tetrahedra with every ion from one trillium lattice connected to the sites of the nearest equilateral triangle from the second trillium lattice (Figure~\ref{fig:structure}b), and therefore we call it tetra-trillium lattice. In this lattice, half of all spins (i.e., those from one trillium lattice) are shared by three tetrahedra, while the remaining half (from the other trillium lattice) belongs to only one tetrahedron. Calculations performed in this limit do not show any signs of magnetic ordering in both classical $S\to\infty$ (down to $T = 0.001 J_4$) and extreme quantum $S=1/2$, 1 limits (down to $\Lambda = 0.01 J_4$) (see supplement), providing strong indications of spin liquid behaviour in both the classical and quantum limits.

The extent of the 'island of liquidity' around the tetra-trillium lattice (indicated by the orange star in Fig.~\ref{fig:TT}a), however, depends strongly on the spin value considered. For $S = 1$, PFFRG indicates a very narrow range around $J_3 = 0$, with an extended set of values along the $J_5/J_4$ axis exhibiting a dynamical ground state, ranging from $\sim 0.8$ up to $\sim 3.5$. On the other hand, the $S = 1/2$ case exhibits the absence of order for $J_3 \neq 0$ for certain values of $J_5/J_4$ and extends up to the sole trillium lattice limit $J_3 = J_4 = 0$. The wider area indicated by the line pattern in Figure~\ref{fig:TT}a indicates a region where it is numerically hard to determine the existence of a breakdown in the $\Lambda$-flow for the $S=1/2$ case. Even beyond this region, the flow breakdowns are quite subtle, indicating weakly ordered phases. The set of parameters calculated with DFT for K$_2$Ni$_2$(SO$_4$)$_3$, indicated by a green star in Fig.~\ref{fig:TT}a, lies close to these paramagnetic regions and many of the features observed in the spin structure factors can be traced back to the tetra-trillium lattice limit.

We also observe the quantum-to-classical correspondence phenomenon for the tetra-trillium lattice, for example for $\Lambda = 0.7~J_4$ in PFFRG and $T=0.3~J_4$ in cMC as can be seen in the bottom part of Fig.~\ref{fig:TT}b. However, as the renormalization group flow continues towards $\Lambda=0$ without any breakdown, the ground state changes giving place to a new structure of correlations (see right panels in Fig.~\ref{fig:TT}b), and also deviates substantially from the cMC calculations at $T=0.001~J_4$, as displayed in the upper part of Figure~\ref{fig:TT}b. This indicates the purely quantum origin of the ground state, i.e., quantum fluctuations can no longer be reproduced by thermal fluctuations in the classical model. The quantum-to-classical correspondence has already been observed to find a limit of validity when the temperature of the quantum solution is lowered and quantum fluctuations start playing a stronger role, even when trying to change the set of parameters from quantum to classical models~\cite{Tao20}.

In the classical limit, cMC calculations show no ordering tendencies in a large $J_5/J_4$ region around the tetra-trillium lattice when $J_3 = 0$. For $J_5<J_4$, a transition to an ordered state occurs within a range $0.3 < J_5/J_4 < 0.4$. For $J_5 > J_4$, no ordering is observed at least up to $J_5/J_4 =3.5$. On the other hand, even the smallest non-zero value of $J_3$ results in a finite critical temperature. This fine line of pure spin-liquid behavior can be understood from the Hamiltonian containing only $J_4$ and $J_5$, which can be written as a disjoint sum of squared total spin in each tetrahedron (just as for the pyrochlore lattice). With this re-writing, the classical ground-state energy can be found exactly given that a solution with zero net magnetic moment in each tetrahedron exists. This solution cannot exist for $J_5/J_4 < 1/3$, providing a theoretical limit (in agreement with cMC calculations). Furthermore, for the tetra-trillium lattice, we confirmed that the solution exists and clusters of 18 spins can be systematically found and flipped while conserving the property of vanishing total spin in each tetrahedron. This allows the system to move through an extensive ground-state manifold and confirms the liquidity of this phase. Compared to the pyrochlore lattice, more spins are needed to construct a minimal flippable cluster because of the ramifications that occur at sites shared by three tetrahedra in the tetra-trillium lattice. By adding $J_1$, $J_2$ or $J_3$, perfect squares cannot be completed, and therefore the freedom to fluctuate is restricted.

The analogy between the tetra-trillium and the pyrochlore lattices can be advanced even further. As demonstrated in Figure~\ref{fig:TT}c, the calculations of $2e(T)$ for a single tetrahedron~\cite{2eorigin} agree quite well with $e(T)$ obtained for all lattice sizes. Only small differences at intermediate temperatures $0.1<T<1.0$ are observed, implying that the tetra-trillium lattice does not behave simply as a set of decoupled tetrahedra. Additionally, the specific heat per site reaches the value $c_v(T\to 0) = 0.75$ as in the pyrochlore lattice and indicates the absence of quartic modes associated with magnetic order~\cite{Moessner98}. On the other hand, there are also significant differences from the pyrochlore lattice. In the latter, the U(1) (or Coulomb) spin liquid can be characterized by algebraically decaying correlations and an emergent gauge field that leads to pinch-point singularities in the spin structure factor~\cite{Isakov04, Henley05}. As shown in Fig.~\ref{fig:TT}, the spin structure factor of the tetra-trillium lattice calculated with cMC at $T=0.001~J_4$ does not present the characteristic pinch-point singularities. This absence can be related to the non-bipartite character of the tetra-trillium lattice in terms of tetrahedra. While in the pyrochlore lattice tetrahedra can be split into two groups (normally dubbed up and down), in the present lattice this is not possible since some corners are shared by three tetrahedra. In two dimensions it has been shown that this can lead to a so-called ${\mathbb{Z}}_{2}$ classical spin liquid, characterized by exponentially decaying correlations down to $T=0$ and fractionalized magnetic moments~\cite{Rehn17} (the properties derive actually from the eigenvalues of the adjacency matrix, which are gapped).

The evidence for QSL behavior in K$_2$Ni$_2$(SO$_4$)$_3$ opens a window of possibilities in search of exotic quantum phases born out of complex 3D lattice geometries beyond the iconic pyrochlore and hyperkagome lattices. Viewing the geometry as a tessellation of tetrahedra on a trillium lattice we have unveiled a so-far unexplored frustration mechanism at play in the langbeinite family. The excellent agreement between experiment and theory demonstrated for K$_2$Ni$_2$(SO$_4$)$_3$ offers an opportunity to further test the applicability of theoretical concepts to a wider variety of compounds that belong to the langbeinite family. Our theoretical phase diagram identifying an `island of liquidity' centered around a highly frustrated tetra-trillium lattice provides a valuable guide in search of further promising QSL candidate materials. It is worth mentioning that despite the appearance of magnetic order outside of the identified regime, the ground states remain highly dynamic thereby allowing for a wide temperature range where emergent phenomena arising out of an interplay between thermal and quantum fluctuations could be explored. An exciting task for future theoretical studies will be to identify possible structures of emergent gauge theories on the tetra-trillium lattice that could underlie the observed QSL behavior. Of particular interest in this context will be how the noncentrosymmetric character of the P2$_1$3 ($\#198$) crystallographic space group of K$_2$Ni$_2$(SO$_4$)$_3$ could give rise to a chiral QSL. 

{\it Methods}

\textbf{Density functional theory based energy mapping}

We determine the Heisenberg Hamiltonian parameters for the $T=100$\,K structure of \ce{K2Ni2(SO4)3} by performing density functional theory (DFT) based energy mapping~\cite{Iqbal2017,Fujihala2022} in the same way as was done for the room temperature structure~\cite{Zivkovic21}. We use all electron DFT calculations with the full potential local orbital (FPLO) basis~\cite{Koepernik1999} and a generalized gradient approximation (GGA) exchange-correlation functional~\cite{Perdew1996}. We correct for strong electronic correlations on the Ni$^{2+}$ $3d$ orbitals using a GGA+U functional~\cite{Liechtenstein1995}. We determine the parameters of the Heisenberg Hamiltonian written in the form
\begin{equation}
    H=\sum_{i<j} J_{ij} {\bf S}_i\cdot {\bf S}_j \,,
\label{eq:hamiltonian}\end{equation}
where ${\bf S}_i$ and ${\bf S}_j$ are spin operators and every bond is counted once.
We create a $\sqrt{2}\times \sqrt{2}\times 1$ supercell with $P2_1$ space group that allows for eight symmetry inequivalent spins. This provides 38 distinct energies of different spin configurations and allows us to resolve the eight nearest neighbour exchange interactions which we name $J_1$ to $J_8$. We choose the relevant value of the interaction $U$ by demanding that the set of interactions match the experimental Curie-Weiss temperature.

\textbf{Inelastic neutron scattering}
Single crystal inelastic neutron scattering data was obtained on the time-of-flight (TOF) instrument LET~\cite{Bewley2011}, ISIS (Didcot, UK) using four different crystals grown from the melt~\cite{Zivkovic21}. The crystals, amounting to a total mass of 0.45g, were co-aligned on aluminium posts, to access the ($HHL$) scattering plane. Cadmium shielding was used to reduce the background signal arising from brass and aluminium. The INS measurements were performed at $T$\,=\,2\,K, with three different incident energies of $E_i$ = 2.8, 5 and 11.7\,meV. The ($HHL$) reciprocal space maps were obtained by scanning over an angular range of 140$^{\circ}$ 
in 1$^{\circ}$ steps for 18\,h, spending 7.5min/$^{\circ}$ for a total of 140 runs. The Horace software was used for visualising and analyzing the four-dimensional $S(\mathbf{Q},\omega)$ data from the TOF experiment \cite{EWINGS2016132}. The single crystal TOF data was symmetrized 
with respect to symmetry operations in the $P2_13$ space group of K$_2$Ni$_2$(SO$_4$)$_3$ ~\cite{Villars2023:sm_isp_sd_0383042,hikita1977phase}. 
In our analysis, we 
normalized and added together data from different equivalent planes based on the space group symmetries. For (100) and (110) family of planes we applied 8 symmetry operations to include reflection and inversion symmetries along the two perpendicular axes and about the origin respectively:(x,y,z), (-x,y,z), (x,-y,z), (x,y,-z), (-x,-y,z), (x,-y,-z), (-x,y,-z), (-x,-y,-z). For (111) family of planes we applied a 6-fold rotational symmetry.

\textbf{Classical Monte Carlo} The cMC calculations were carried out using a logarithmic cooling protocol from $T=2~J_4$ down to $T=0.001~J_4$ with 150 temperature steps. Systems of up to $8\times 12^3= 13824$ unitary spins are considered. At each temperature, $10^5$ cMC steps were performed. Each cMC step consists of N Metropolis trials and N overrelaxation steps intercalated. The acceptance rate of the Metropolis trials is kept at $50\%$ using the adapted Gaussian step~\cite{Alzate19}. Correlations are calculated over already thermalized states at selected temperatures by doing $4\times 10^5$ cMC steps while measuring correlations once every 100 steps. All results are then averaged over 5 independent runs. The spin structure factors are calculated taking into account the positions of the Ni atoms corresponding to each model, as well as the form factor corresponding to Ni atoms.

\textbf{Pseudo-fermion functional renormalization group}
All results for $S=1/2$ and $S=1$ quantum spin models in this paper are obtained by standard PFFRG \cite{muller2023pseudofermion} with the following specifications (further information on the method is provided in the supplementary material).
Vertex frequency dependencies are approximated by finite grids with exponentially distributed frequencies. The self-energy is evaluated for $1000$ positive frequency arguments. Frequency grids for the two-particle vertex contain $32$ positive values for each of the three transfer frequencies. Spin correlations spanning over distances larger than three lattice constants of the underlying cubic lattice are neglected. After consideration of lattice translation symmetries, this implies the computation of spin-correlations along $1842$ lattice vectors, or $622$ vectors unrelated by lattice symmetries. For the computation of the phase diagram Fig. \ref{fig:TT}a, the maximum included correlation vector distance is reduced to two lattice constants, implying $186$ vectors unrelated by symmetry. 
Flow equations are solved by application of an explicit embedded Runge-Kutta (2, 3) method with adaptive step size \cite{GSL09}.

{\it Acknowledgments}. M.G.G. and V.N. would like to thank the HPC Service of ZEDAT, Freie Universität Berlin, for computing time. V.N. gratefully acknowledges the computing time provided on the high-performance computers of Noctua 2 at the NHR Center PC2. These are funded by the Federal Ministry of Education and Research and the state governments participating on the basis of the resolutions of the GWK for the national highperformance computing at universities (www.nhr-verein.de/unsere-partner). Furthermore, V.N. would like to thank the Tron cluster service at the Department of Physics, Freie Universität Berlin for the computing time provided. A.S. acknowledges Swiss government FCS grant(2021.0414) for providing support during the work. The work of Y.I. was performed in part at the Aspen Center for Physics, which is supported by National Science Foundation grant PHY-2210452. The participation of Y.I. at the Aspen Center for Physics was supported by the Simons Foundation. This research was supported in part by the National Science Foundation under Grant No. NSF PHY-1748958. Y.I. acknowledges support from the ICTP through the Associates Programme and from the Simons Foundation through grant number 284558FY19, IIT Madras through QuCenDiEM (Project No. SP22231244CPETWOQCDHOC), the International Centre for Theoretical Sciences (ICTS), Bengaluru, India during a visit for participating in the program “Frustrated Metals and Insulators” (Code: ICTS/frumi2022/9). Y.I. acknowledges the use of the computing resources at HPCE, IIT Madras. J. R. and H.O.J. thank IIT Madras for a Visiting Faculty Fellow position under the IoE program which facilitated the completion of this work and writing of the manuscript. H.M.R. acknowledges the funding by the European Research Council (ERC) under the European Union’s Horizon 2020 research and innovation program projects HERO (Grant No. 810451) and Swiss National Science Foundation (SNF) Quantum Magnetism grant (No. 200020-188648). Inelastic neutron data has been obtained at the LET beamline, ISIS (RB1910466).

{\it Author contributions}. Single crystals were grown by A.M. Inelastic neutron scattering data was measured and analyzed by A.S., V.F., J-R. S., R.B., H.M.R. and I.\v{Z}. Classical Monte Carlo calculations were carried out by M.G.G and J.R. Pseudo-fermion functional renormalization group calculations were carried out by V.N. and J.R. Energy mappings based on density-functional theory were carried out by H.O.J. Analytical calculations on the tetra-trillium lattice were carried out by M.G.G, V.N., J.R., and Y.I. The manuscript was written by M.G.G. and I.\v{Z}. with contributions from all co-authors.

\bibliography{papers}

\begin{thebibliography}{55}
\bibitem [{2eo()}]{2eorigin}%
  \BibitemOpen
  \href@noop {} {}\bibinfo {note} {In the tetra-trillium lattice, on average,
  each site is shared by 2 tetrahedra. This means that there are 2 sites per
  tetrahedron in the system. On the other hand, on the isolated tetrahedron,
  there are 4 sites. Therefore, a factor 2 has to be added in the energy per
  site to compensate.}\BibitemShut {Stop}%
\end{thebibliography}

\end{document}




\title{Supplementary Information for "Dynamics of K$_2$Ni$_2$(SO$_4$)$_3$ governed by proximity to a 3D spin liquid model"}

\author{M. G. Gonzalez}
\affiliation{Helmholtz-Zentrum Berlin f\"ur Materialien und Energie, Hahn-Meitner-Platz 1, 14109 Berlin, Germany}
\affiliation{Dahlem Center for Complex Quantum Systems and Fachbereich Physik, Freie Universität Berlin, 14195 Berlin, Germany\looseness=-1}

\author{V. Noculak}
\affiliation{Helmholtz-Zentrum Berlin f\"ur Materialien und Energie, Hahn-Meitner-Platz 1, 14109 Berlin, Germany}
\affiliation{Dahlem Center for Complex Quantum Systems and Fachbereich Physik, Freie Universität Berlin, 14195 Berlin, Germany\looseness=-1}

\author{A. Sharma}
\affiliation{Laboratory for Quantum Magnetism, Institute of Physics, \'Ecole Polytechnique F\'ed\'erale de Lausanne, CH-1015 Lausanne, Switzerland\looseness=-1}

\author{V. Favre}
\affiliation{Laboratory for Quantum Magnetism, Institute of Physics, \'Ecole Polytechnique F\'ed\'erale de Lausanne, CH-1015 Lausanne, Switzerland\looseness=-1}

\author{J-R. Soh}
\affiliation{Laboratory for Quantum Magnetism, Institute of Physics, \'Ecole Polytechnique F\'ed\'erale de Lausanne, CH-1015 Lausanne, Switzerland\looseness=-1}

\author{A. Magrez}
\affiliation{Crystal Growth Facility, \'Ecole Polytechnique Fédérale de Lausanne, Lausanne, Switzerland.\looseness=-1}

\author{R. Bewley}
\affiliation{ISIS Pulsed Neutron and Muon Source, STFC Rutherford Appleton Laboratory, Harwell Science and Innovation Campus, Didcot, Oxfordshire, OX11 0QX UK\looseness=-1}

\author{Harald O. Jeschke}
\affiliation{Research Institute for Interdisciplinary Science, Okayama University, Okayama 700-8530, Japan}

\author{J. Reuther}
\affiliation{Helmholtz-Zentrum Berlin f\"ur Materialien und Energie, Hahn-Meitner-Platz 1, 14109 Berlin, Germany}
\affiliation{Dahlem Center for Complex Quantum Systems and Fachbereich Physik, Freie Universität Berlin, 14195 Berlin, Germany\looseness=-1}
\affiliation{Department of Physics and Quantum Centers in Diamond and Emerging Materials (QuCenDiEM) Group,
Indian Institute of Technology Madras, Chennai 600036, India\looseness=-1}

\author{H. M. R{\o}nnow}
\affiliation{Laboratory for Quantum Magnetism, Institute of Physics, \'Ecole Polytechnique F\'ed\'erale de Lausanne, CH-1015 Lausanne, Switzerland\looseness=-1}

\author{Y. Iqbal}
\affiliation{Department of Physics and Quantum Centers in Diamond and Emerging Materials (QuCenDiEM) Group,
Indian Institute of Technology Madras, Chennai 600036, India\looseness=-1}

\author{I. \v{Z}ivkovi\'c}
\affiliation{Laboratory for Quantum Magnetism, Institute of Physics, \'Ecole Polytechnique F\'ed\'erale de Lausanne, CH-1015 Lausanne, Switzerland\looseness=-1}

\maketitle
\beginsupplement

\onecolumngrid


\section{DFT energy mapping for the low-temperature structure}

\begin{figure}[!h]
\centering
\includegraphics[width=0.4\textwidth]{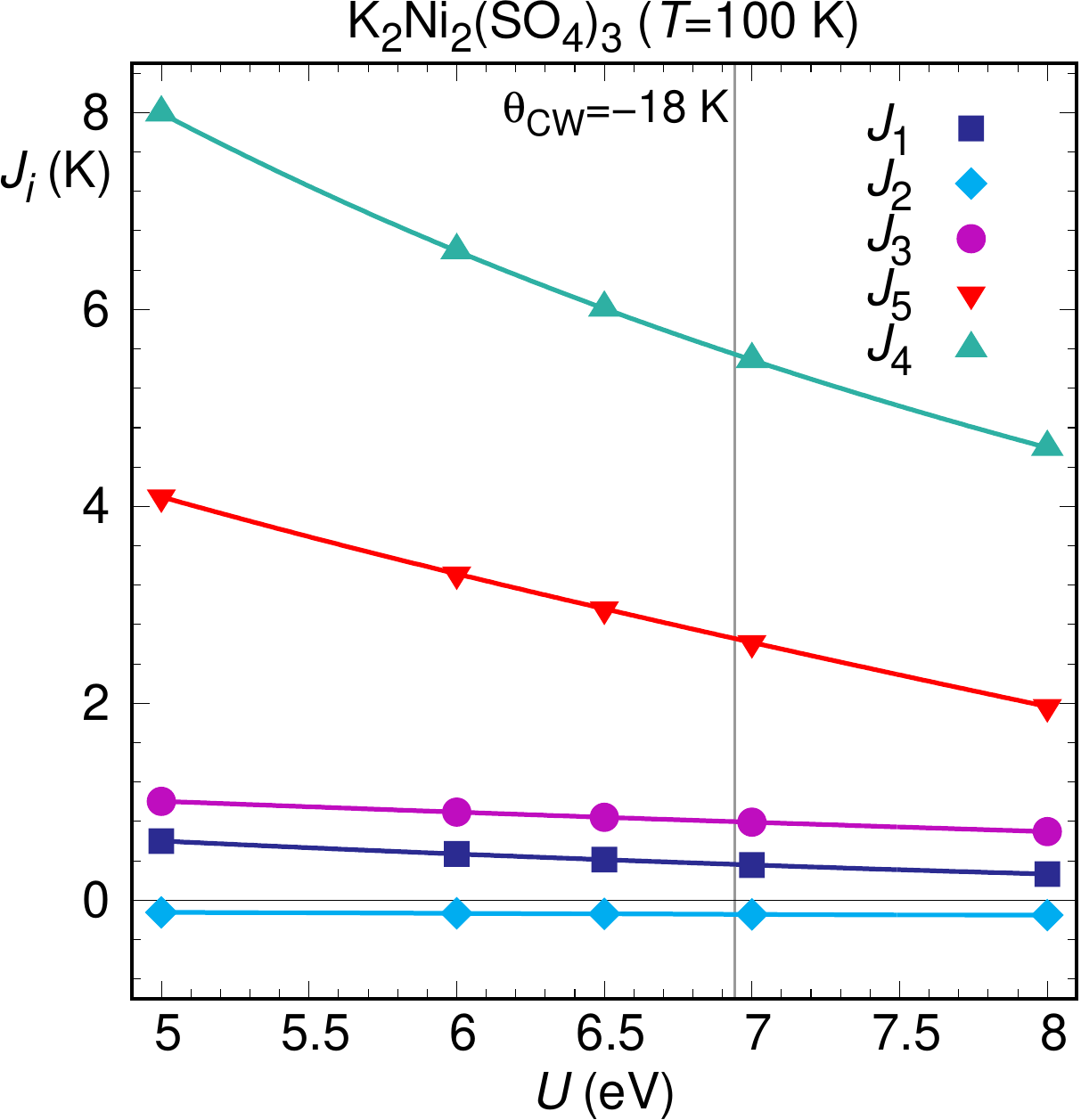}
\caption{DFT energy mapping for the $T=100$\,K structure of {\kni} calculated with DFT+U as function of onsite interaction strength $U$. The Hund's rule coupling value is fixed at $J_{\rm H}=0.88$\,eV~\cite{Mizokawa1996}. The vertical line indicates the $U$}
\label{fig:100Kcouplings}
\end{figure}

\begin{table*}[htb]
    \begin{tabular}{c|c|c|c|c|c|c|c|c|c}
    $U$\,(eV)&$J_1$\,(K)&$J_3$\,(K)&$J_2$\,(K)&$J_5$\,(K)&$J_4$\,(K)&$J_6$\,(K)&$J_7$\,(K)&$J_8$\,(K)&$\theta_{CW}$\,(K)\\\hline
5 & 0.602(2) & -0.123(1) & 1.005(2) & 4.097(1) & 7.992(1) & 0.025(1) & -0.012(1) & 0.022(1) & -26.4 \\
6 & 0.470(2) & -0.133(1) & 0.895(1) & 3.315(1) & 6.594(1) & 0.020(1) & -0.009(1) & 0.018(1) & -21.7 \\
6.5 & 0.412(2) & -0.139(1) & 0.842(2) & 2.960(1) & 6.009(1) & 0.0169(1) & -0.008(1) & 0.016(1) & -19.7 \\
{\bf 6.942} & {\bf 0.364(2)} & {\bf -0.144(1)} & {\bf 0.798(2)} & {\bf 2.657(1)} & {\bf 5.545(1)} & {\bf 0.015(1)} & {\bf -0.007(1)} & {\bf 0.014(1)} & {\bf -18} \\
7 & 0.358(2) & -0.144(1) & 0.792(2) & 2.618(1) & 5.487(1) & 0.015(1) & -0.007(1) & 0.014(1) & -17.8 \\
7.5 & 0.312(1) & -0.148(1) & 0.744(1) & 2.289(1) & 5.017(1) & 0.012(1) & -0.007(1) & 0.013(1) & -16.1 \\
8 & 0.266(1) & -0.152(1) & 0.696(1) & 1.967(1) & 4.596(1) & 0.010(1) & -0.005(1) & 0.011(1) & -14.4 \\\hline
$d_{\rm Ni-Ni}$\,{\AA} & 4.405 & 4.889 & 6.072 & 6.112 & 6.114 & 8.116 & 8.387 & 8.563 & 
    \end{tabular}
    \caption{Exchange interactions of {\kni}, calculated for six different values of the on-site interaction strength $U$. The line in bold face is interpolated to yield the experimental value of the Curie-Weiss temperature (Ref.~\cite{Zivkovic21}), using Eq.~\eqref{eq:tcw}. Statistical errors from the fitting procedure are given.}
    \label{tab:couplings}
\end{table*}

 Fig.~\ref{fig:100Kcouplings} shows the result of the energy mapping procedure for 6 different values of the onsite interaction strength $U$; the Hund's rule coupling was fixed to $J_{\rm H}=0.88$\,eV following Ref.~\cite{Mizokawa1996}. The long-range couplings $J_6$ to $J_8$ turn out to be very close to zero; they are listed in Table~\ref{tab:couplings} but not shown in Fig.~\ref{fig:100Kcouplings}. Note that resolving these small long-range couplings is not irrelevant because on the one hand, we find proof that they do not play a role as opposed to just assuming this is the case. On the other hand, resolving fewer exchange interactions in the energy mapping procedure has the consequence that long-range couplings are added into short-range couplings with some prefactor -- this is only harmless if the long-range couplings are really effectively zero. Note that between room temperature and $T=100$\,K structure, $J_4$ and $J_5$ would switch places if we strictly number exchange couplings by Ni-Ni distance. Therefore, as can be seen in Table~\ref{tab:couplings}, we use the name $J_5$ for the second trillium coupling and $J_4$ for the dominant coupling connecting trillium lattices also in the low-temperature structure. In this way, we can see that cooling the lattice of {\kni} leads to moderate but, as we demonstrate in the main text, significant adjustments in the Heisenberg Hamiltonian parameters.
We choose the relevant value of the interaction $U$ by demanding that the set of interactions match the experimental Curie-Weiss temperature of $\theta_{\rm CW}=-18$~K.
From the Heisenberg exchange interactions, we determine the Curie-Weiss temperature according to
\begin{equation}
       \theta_{\rm CW}=-\frac{1}{3}S(S+1)\big( J_1 + 3J_2 + 3 J_3 + 3 J_4 + 3 J_5 + 3 J_6 + 3 J_7 + 3 J_8 \big) \,,
\label{eq:tcw}\end{equation}
where $S=1$.

\section{cMC thermodynamics}

We perform classical Monte Carlo calculations on systems of $N=8\times L^3$ classical unitary spins, with different sizes $L$ from 2 ($N=64$) and up to 12 ($N=13824$). Generally, a logarithmic cool down with 150 steps is applied from $T=2.0~J_4$ down to $0.001~J_4$ (a finer grid can be applied later to any particular region of interest, e. g. a phase transition). At each temperature, $10^5$ Monte Carlo steps are applied, each consisting of $N$ single-site Metropolis trials and $N$ over-relaxation updates (intercalated). We also fix the acceptance ratio of the Metropolis trial close to $50\%$ by using the so-called Gaussian step~\cite{Alzate19}. Data for the calculation of the energy $e$ and specific heat $c_v$ are collected during the second half of the Monte Carlo steps at each temperature and later averaged over 5 independent runs. 

\begin{figure}[!h]
\centering
\includegraphics[width=0.97\textwidth]{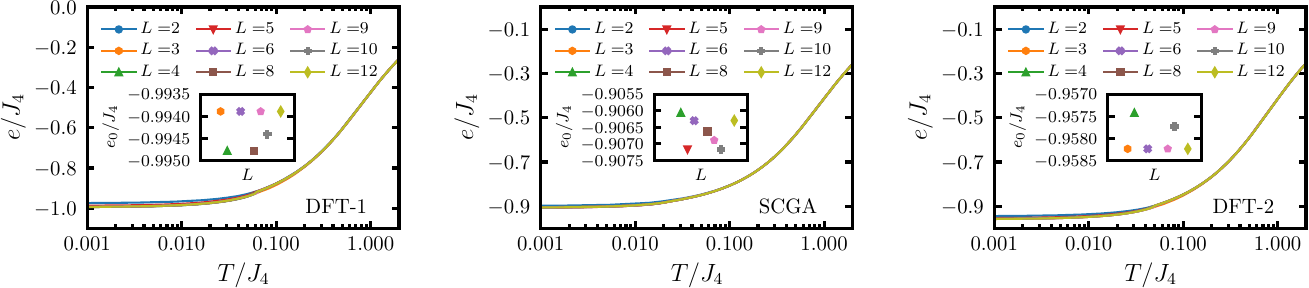}
\caption{Energy as a function of $T/J_4$ for three different models of K$_2$Ni$_2$(SO$_4$)$_3$: DFT-1~\cite{Zivkovic21}, SCGA~\cite{Yao23}, and DFT-2 corresponding to the updated structure and parameters presented in this article. Insets show the ground-state energy obtained in the $T\to 0$ limit, depending on the lattice size $L$.}
\label{fig:cmcres}
\end{figure}

Figure~\ref{fig:cmcres} shows the results of $e/J_4(T)$ for the three different models: the DFT model from Ref.~\cite{Zivkovic21} (DFT-1, left), the SCGA model from Ref.~\cite{Yao23} (middle), and the new DFT model presented in this article taking into account the 100~K structure (DFT-2, right). Phase transitions can be seen as very small features in each curve, showing the highly-frustrated nature of the models. The inset in each panel shows the ground-state energy obtained in the limit $T\to 0$ as a function of the size $L$. For DFT-1, the lowest energy is obtained for the $L=4n$ lattices. However, $L=12$ converges to the $L=3n$ energy, a little above. Since $L=12$ can fit the $L=4$ solution which has lower energy, a transition should occur from one state to the other. However, orders are complicated and different from one another, so that $L=12$ gets trapped in a stable excited state. For SCGA, the lowest energy is obtained for $L=5n$, and the $L=3n$ observed experimentally is not observed. For the new DFT-2 model, the lowest energy is found for $L=3n$. In this case, $L=8$ gets trapped in a higher energy state, but it should be in the same line as $L=4$ at least.

\section{PFFRG flows}

The PFFRG method~\cite{Reuther10} relies on writing the quantum spin operators in terms of $S=1/2$ fermionic operators. Using many-body Feynman diagrammatic techniques and introducing an infrared frequency cutoff $\Lambda$ leads to an infinite set of coupled differential equations for the corresponding fermionic vertex functions, from which the spin-spin correlations at $T=0$ can be obtained by choosing an appropriate truncation scheme. In this case, we use the one-loop approximation to resolve the flow in real space, starting from known initial conditions at $\Lambda \to \infty$. Most importantly, PFFRG contains the correct large-$N$ (spin flavour) and large-$S$ (spin size) limits, making it ideal to study quantum systems. Higher values of $S>1/2$ are obtained coupling copies of the $S=1/2$ fermionic spins at each site~\cite{Baez17} (in the present case for $S=1$ only two are needed).

In particular, since PFFRG preserves all symmetries of the original Hamiltonian, the onset of magnetic order (symmetry breaking) is signalled by a breakdown in the $\Lambda$ flow. This means that a peak or kink appears in the susceptibility as a function of $\Lambda$. On the other hand, if there is no magnetic order (all symmetries are preserved), the flow continues smoothly down to $\Lambda \ll J_4$. We also exploit the translation symmetry of the lattice and calculate spin-spin correlations only up to a certain distance from the reference sites (those not connected by lattice symmetries). In this case, we take distances up to 3 unit cells, implying 1842 lattice vectors that can be reduced to 622 after exploiting the lattice symmetries. 

\begin{figure}[!h]
\centering
\includegraphics[width=0.4\textwidth]{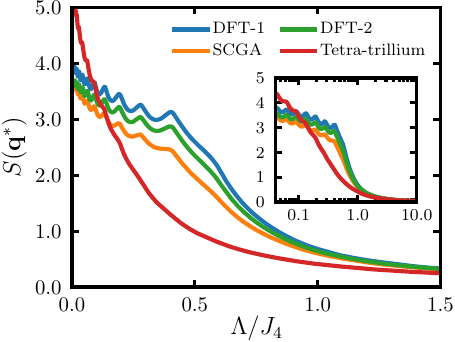}
\caption{PFFRG flows for different models. For each case, $\mathbf{q}^*$ corresponds to the point in reciprocal space for which the highest value of the spin structure factor is observed.}
\label{fig:flows}
\end{figure}

Figure~\ref{fig:flows} shows examples of $\Lambda$-flows for the DFT-1,  DFT-2, SCGA, and tetra-trillium models. For the first two, a breakdown is clearly observed, indicating some kind of symmetry breaking in the ground state of the corresponding models. On the other hand, the flow of the tetra-trillium lattice continues to the lowest values of $\Lambda$ without presenting any peaks or kinks, indicating no symmetry breaking in this case. For each case, $\mathbf{q}^*$ corresponds to the point in reciprocal space for which the highest value of the spin structure factor is observed in the low cutoff limit (also for the flows with breakdown).

\section{Spin structure factor for other models}

\begin{figure}[!h]
\centering
\includegraphics[width=0.97\textwidth]{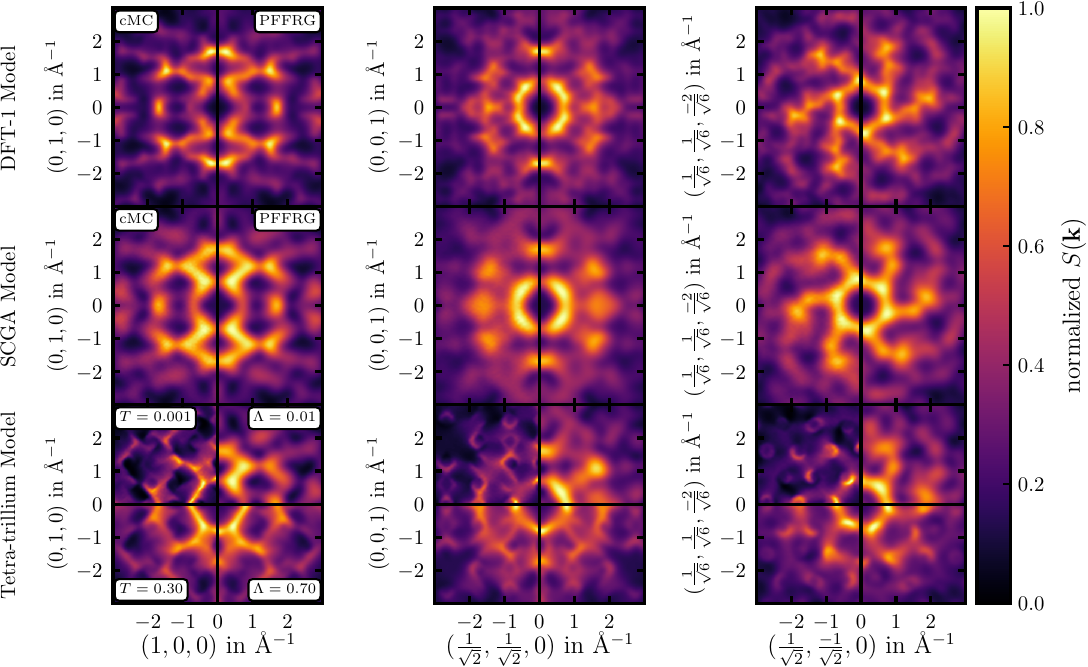}
\caption{Spin structure factor obtained with cMC (left parts) and PFFRG (right parts). The first two lines correspond to DFT-1~\cite{Zivkovic21}, DFT-2 (this article), and SCGA models~\cite{Yao23}; and calculations are at $T=0.3 J_4$ for cMC and at the breaking point of the flow for PFFRG. The bottom row corresponds to the tetra-trillium lattice, and temperatures and $\Lambda$-cutoffs are indicated in the panels.}
\label{fig:ssf}
\end{figure}

The cMC data for the correlations are calculated separately from an already thermalized configuration at a given temperature $T$, by doing $4\times 10^5$ Monte Carlo steps while measuring once every $10^2$, averaging over 5 independent runs. The results are shown in Fig.~\ref{fig:ssf}, along with the PFFRG calculations. The first two rows correspond to the DFT-1 and SCGA models~\cite{Zivkovic21, Yao23}, respectively. For these, cMC calculations are performed at $T=0.3~J_4$ and PFFRG calculations are performed at the point where the $\Lambda$-flow breaks. The quantum-to-classical correspondence phenomenon is observed in both cases. The last row corresponds to the tetra-trillium lattice, where there is no breakdown of the flow in PFFRG. Therefore, correlations can be calculated at the lowest $\Lambda = 0.01 J_4$. When these results are compared against cMC calculations at low temperatures, there is no agreement between the patterns. However, quantum-to-classical correspondence is recovered at finite $T$ and $\Lambda$, as shown in the lower parts of the last row of panels.

\section{Ground state of the tetra-trillium lattice}

\begin{figure}[!h]
\centering
\includegraphics[width=0.97\textwidth]{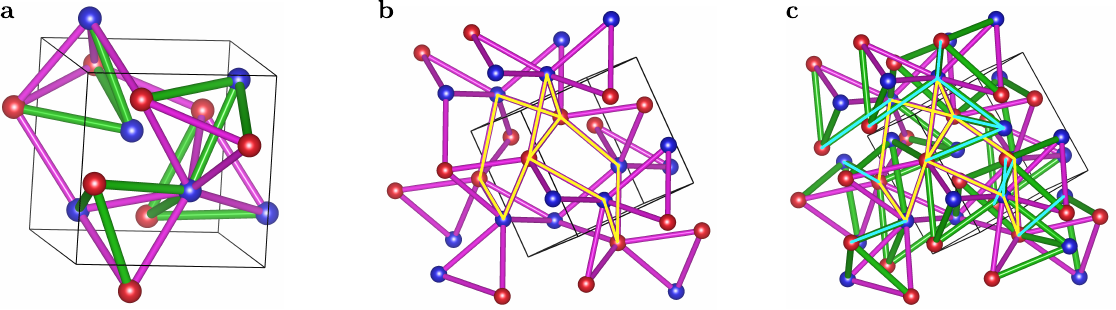}
\caption{(a) one possible ground-state configuration defined in one unit cell. (b) flippable loop on the $J_5$ trillium lattice. (c) flippable loop on the tetra-trillium lattice.}
\label{fig:loops}
\end{figure}

As mentioned in the main article, classical ground-state spin configurations for the tetra-trillium lattice can be easily found. In Fig.~\ref{fig:loops}a, a simple configuration is shown where blue and red sites correspond to two opposite directions for spins (up and down). In this particular configuration, half of the spins in each lattice point up, and the remaining half point down. Thus, both sublattices have total zero magnetization. We have verified numerically that the configuration defined on a single unit cell can be copied throughout the whole lattice and the ground-state energy is obtained. We show in Fig.~\ref{fig:loops}b a larger part of the lattice, but only taking into account the $J_5$ trillium sublattice lattice. In this sublattice, a loop of 9 spins can be defined as shown by the yellow paths, consisting of two pentagons and one triangle that share a common base. Flipping the 9 spins in this loop leaves all triangles in the trillium sublattice with two spins pointing in one direction and the third in the other. This means, no triangles are left with three spins pointing in the same direction. Therefore, the zero-sum in each tetrahedron can be satisfied by choosing accordingly the spin on the $J_3$ trillium sublattice. This is shown in Fig.~\ref{fig:loops}c, where the remaining 9 spins are connected to the yellow loop by light-blue lines. This total of 18 spins can be flipped while maintaining the ground-state energy. We have verified that these loops can be found systematically departing from any site. A detailed explanation will be presented elsewhere.

\bibliography{papers}